\documentclass[
    ,final            
  ]
  {aipproc}

\layoutstyle{6x9}

\begin{document}

\def\jnl@style{\it}
\def\aaref@jnl#1{{\jnl@style#1}}

\def\aaref@jnl#1{{\jnl@style#1}}

\def\aj{\aaref@jnl{AJ}}                   
\def\araa{\aaref@jnl{ARA\&A}}             
\def\apj{\aaref@jnl{ApJ}}                 
\def\apjl{\aaref@jnl{ApJ}}                
\def\apjs{\aaref@jnl{ApJS}}               
\def\ao{\aaref@jnl{Appl.~Opt.}}           
\def\apss{\aaref@jnl{Ap\&SS}}             
\def\aap{\aaref@jnl{A\&A}}                
\def\aapr{\aaref@jnl{A\&A~Rev.}}          
\def\aaps{\aaref@jnl{A\&AS}}              
\def\azh{\aaref@jnl{AZh}}                 
\def\baas{\aaref@jnl{BAAS}}               
\def\jrasc{\aaref@jnl{JRASC}}             
\def\memras{\aaref@jnl{MmRAS}}            
\def\mnras{\aaref@jnl{MNRAS}}             
\def\pra{\aaref@jnl{Phys.~Rev.~A}}        
\def\prb{\aaref@jnl{Phys.~Rev.~B}}        
\def\prc{\aaref@jnl{Phys.~Rev.~C}}        
\def\prd{\aaref@jnl{Phys.~Rev.~D}}        
\def\pre{\aaref@jnl{Phys.~Rev.~E}}        
\def\prl{\aaref@jnl{Phys.~Rev.~Lett.}}    
\def\pasp{\aaref@jnl{PASP}}               
\def\pasj{\aaref@jnl{PASJ}}               
\def\pasa{\aaref@jnl{PASA}}               
\def\qjras{\aaref@jnl{QJRAS}}             
\def\skytel{\aaref@jnl{S\&T}}             
\def\solphys{\aaref@jnl{Sol.~Phys.}}      
\def\sovast{\aaref@jnl{Soviet~Ast.}}      
\def\ssr{\aaref@jnl{Space~Sci.~Rev.}}     
\def\zap{\aaref@jnl{ZAp}}                 
\def\nat{\aaref@jnl{Nature}}              
\def\science{\aaref@jnl{Science}}              
\def\iaucirc{\aaref@jnl{IAU~Circ.}}       
\def\aplett{\aaref@jnl{Astrophys.~Lett.}} 
\def\apspr{\aaref@jnl{Astrophys.~Space~Phys.~Res.}}
\def\bain{\aaref@jnl{Bull.~Astron.~Inst.~Netherlands}} 
\def\fcp{\aaref@jnl{Fund.~Cosmic~Phys.}}  
\def\gca{\aaref@jnl{Geochim.~Cosmochim.~Acta}}   
\def\grl{\aaref@jnl{Geophys.~Res.~Lett.}} 
\def\jcp{\aaref@jnl{J.~Chem.~Phys.}}      
\def\jgr{\aaref@jnl{J.~Geophys.~Res.}}    
\def\jqsrt{\aaref@jnl{J.~Quant.~Spec.~Radiat.~Transf.}}
\def\memsai{\aaref@jnl{Mem.~Soc.~Astron.~Italiana}}
\def\nphysa{\aaref@jnl{Nucl.~Phys.~A}}   
\def\physrep{\aaref@jnl{Phys.~Rep.}}   
\def\physscr{\aaref@jnl{Phys.~Scr}}   
\def\planss{\aaref@jnl{Planet.~Space~Sci.}}   
\def\procspie{\aaref@jnl{Proc.~SPIE}}   

\def\crasp{\aaref@jnl{C.~R.~Acad.~Sci.~Paris}}   

\let\astap=\aap
\let\apjlett=\apjl
\let\apjsupp=\apjs
\let\applopt=\ao


\newcommand\eex[1]{\mbox{$\times 10^{#1}$}}              
\newcommand\eez[1]{\mbox{$10^{#1}$}}                     
\newcommand\ion[2]{\mbox{#1$\,${\small\rmfamily{#2}}}}   
\def\lesssim{\mathrel{\hbox{\rlap{\hbox{\lower4pt\hbox{$\sim$}}}\hbox{$<$}}}}
\def\hi{\ion{H}{I}}
\def\hii{\ion{H}{II}}
\def\civ{\ion{C}{IV}}
\def\ovi{\ion{O}{VI}}
\def\sq{\mbox{\rlap{$\sqcap$}$\sqcup$}}%
\def\arcdeg{\mbox{$^\circ$}}%
\def\arcmin{\mbox{$^\prime$}}%
\def\arcsec{\mbox{$^{\prime\prime}$}}%
\def\elec{\mbox{e$^{-}$}}%
\def\raday{\mbox{$^{\mathrm d}$}}%
\def\rahour{\mbox{$^{\mathrm h}$}}%
\def\ramin{\mbox{$^{\mathrm m}$}}%
\def\rasec{\mbox{$^{\mathrm s}$}}%
\def\fday{\mbox{$.\!\!^{\mathrm d}$}}%
\def\fhour{\mbox{$.\!\!^{\mathrm h}$}}%
\def\fmin{\mbox{$.\!\!^{\mathrm m}$}}%
\def\fsec{\mbox{$.\!\!^{\mathrm s}$}}%
\def\farcdeg{\mbox{$.\!\!^\circ$}}%
\def\farcmin{\mbox{$.\!\!^\prime$}}%
\def\farcsec{\mbox{$.\!\!^{\prime\prime}$}}%
\def\halpha{\mbox{H$\alpha$}}
\def\vdev{\mbox{$v_{\rm dev}$}}
\def\vhel{\mbox{$v_{\rm helio}$}}
\def\msun{\hbox{M$_{\odot}$}}
\def\mdot{\hbox{$\dot M$}}
\def\zsun{Z_{\odot}}
\def\rsloan{\hbox{$r^\prime$}}
\def\sqarc{\hbox{arcsec$^{2}$}}
\def\psqarc{\hbox{arcsec$^{-2}$}}
\def\lsun{\hbox{L$_{\odot}$}}
\def\imfa{\hbox{$\delta(m-1)$}}
\def\imfb{\hbox{$m^{-2.35}$}}
\def\rana{\hbox{1}}
\def\ranb{\hbox{[0.2,1]}}
\def\ranc{\hbox{[0.2,100]}}
\def\SN{signal-to-noise ratio}
\def\mass{$M_{\odot}$}
\def\deg{^{\circ}}
\def\kps{\mbox{${\rm km~s^{-1}}$}}
\def\snu{\mbox{$_\nu$}}
\def\tb{\mbox{$T_b$}}
\def\ts{\mbox{$T_s$}}
\def\slambda{\mbox{$S_{\lambda}$}}
\def\cm{\mbox{${\rm cm^{-2}}$}}
\def\nh{\mbox{$N_{\rm HI}$}}
\def\NH{\nh}
\def\mh{\mbox{$M_{\rm HI}$}}
\def\mhi{\mbox{$M_{\rm HI}$}}
\def\mp{\mbox{$m_p$}}
\def\sunits{\mbox{${\rm Jy~beam^{-1}}$}}
\newcommand\R{\mbox{$R$}}
\newcommand\U{\mbox{$U$}}
\newcommand\B{\mbox{$B$}}
\newcommand\V{\mbox{$V$}}
\newcommand\ubvr{\mbox{$U\!BV\!R$}}
\newcommand\ub{\mbox{$U\!-\!B$}}
\newcommand\bv{\mbox{$B\!-\!V$}}
\newcommand\vr{\mbox{$V\!-\!R$}}
\newcommand\ur{\mbox{$U\!-\!R$}}
\newcommand\phn{\phantom{0}}%
\newcommand\phd{\phantom{.}}%
\newcommand\phs{\phantom{$-$}}%
\newcommand\phm[1]{\phantom{#1}}%
\newcommand{\etal}{{\it et al.\/}}
\newcommand{\ie}{{\it i.e.\/}}
\newcommand{\eg}{{\it e.g.\/}}
\newcommand{\nd}{\nodata}
\newcommand{\pv}{\mbox{$p$-$v$}}


\newcommand\bls{\baselineskip}
\newcommand\npb{\nopagebreak}
\newcommand\eqsp{\vspace{2pt}}
\newcommand{\D}{\displaystyle}
\newcommand{\bc}{\begin{center}}
\newcommand{\ec}{\end{center}}
\newcommand{\addc}{\addtocounter}
\newcommand{\setc}{\setcounter}
\newcommand{\acl}{\addcontentsline}
\newcommand{\np}{\newpage}
\newcommand{\tpsp}{\thispagestyle{plain}}

\def\ergss   {erg~s$^{-1}$}
\def\ergscms   {erg~cm$^{-2}$~s$^{-1}$}
\def\ergscmsdeg   {erg~cm$^{-2}$~s$^{-1}$~deg$^{-2}$}
\def\rth   {\mbox{$r_{200}$}}
\def\rfh   {\mbox{$r_{500}$}}
\def\rtfh   {\mbox{$r_{2500}$}}
\def\rvir   {\mbox{$r_{\rm vir}$}}
\def\suzaku   {\textit{Suzaku\/}}
\def\chandra  {\textit{Chandra\/}}
\def\xmm      {\textit{XMM\/}}
\def\xmmnewton      {\textit{XMM-Newton\/}}
\def\rosat      {\textit{ROSAT\/}}
\def\fgas      {\mbox{$f_{\rm gas}$}}

\title{The Outer Limits of Galaxy Clusters: Observations to the Virial
Radius with\\Suzaku, XMM, and Chandra}



\author{Eric D.\ Miller}{
  address={MIT Kavli Institute for Astrophysics and Space Research, 
  Cambridge, MA, USA; milleric@mit.edu}
}
\author{Marshall Bautz}{
  address={MIT Kavli Institute for Astrophysics and Space Research, 
  Cambridge, MA, USA; milleric@mit.edu}
}
\author{Jithin George}{
  address={Astronomy Department, University of Maryland, College Park, MD, USA}
}
\author{Richard Mushotzky}{
  address={Astronomy Department, University of Maryland, College Park, MD, USA}
}
\author{David Davis}{
  address={CRESST and X-ray Astrophysics Laboratory, NASA/GSFC, Greenbelt, MD, USA},altaddress={Department of Physics, University of Maryland Baltimore County, Baltimore, MD, USA}
}
\author{J.\ Patrick Henry}{
  address={Institute for Astronomy, University of Hawaii, Honolulu, HI, USA}
}

\begin{abstract}
The outskirts of galaxy clusters, near the virial radius, remain relatively
unexplored territory and yet are vital to our understanding of cluster
growth, structure, and mass.  In this presentation, we show the first
results from a program to constrain the state of the outer intra-cluster
medium (ICM) in a large sample of galaxy clusters, exploiting the strengths
of three complementary X-ray observatories: \suzaku\ (low, stable
background), \xmmnewton\ (high sensitivity), and \chandra\ (good spatial
resolution).  By carefully combining observations from the cluster core to
beyond \rth, we are able to identify and reduce systematic uncertainties
that would impede our spatial and spectral analysis using a single
telescope.  Our sample comprises nine clusters at $z \sim 0.1$--0.2
fully covered in azimuth to beyond \rth, and our analysis indicates that
the ICM is not in hydrostatic equilibrium in the cluster outskirts, where
we see clear azimuthal variations in temperature and surface brightness.
In one of the clusters, we are able to measure the diffuse X-ray emission
well beyond \rth, and we find that the entropy profile and the gas fraction
are consistent with expectations from theory and numerical simulations.
These results stands in contrast to recent studies which point to gas
clumping in the outskirts; the extent to which differences of cluster
environment or instrumental effects factor in this difference remains
unclear.  From a broader perspective, this project will produce a sizeable
fiducial data set for detailed comparison with high-resolution numerical
simulations.
\end{abstract}

\maketitle


\section{Cluster Outskirts: The Final Frontier}

Exploring the outskirts of galaxy clusters is vital to our understanding of
cluster growth, structure, and mass.  The region near the boundary of the
dynamically relaxed cluster volume ($\sim \rth$ \cite{Evrardetal1996}),
offers a direct view of cluster growth and ICM accretion \cite{Voit2005}.
Simulations predict deviations from hydrostatic equilibrium in the
outskirts of clusters at scales on the order of $\rtfh<r<\rth$, well beyond
the influence of cooling and AGN feedback processes in the  cluster core
\cite{Roncarellietal2006,Burnsetal2010}.  Moreover, it may be that
signatures of spatially inhomogeneous accretion from cosmic filaments are
observable at these radii. X-ray observations of clusters at $r>\rfh$ also
probe the distribution of mass to large radii, and in so doing provide new
constraints on structure formation models
\cite{NFW1997,Borganietal2004,Roncarellietal2006}.  Finally, accurate
knowledge of the mass and baryonic fraction of  clusters is essential for
cluster-based cosmological tests \cite{Vikhlininetal2009a,Allenetal2004}.
Accurate cluster masses within \rfh, for example, required for many
cosmological tests, rely on knowledge of both the temperature and its
radial gradient at \rfh, and observations to larger radius are essential to
determine these quantities reliably. 

\begin{figure}[t]
\begin{minipage}{.38\textwidth}
\includegraphics[height=\textwidth,angle=270]{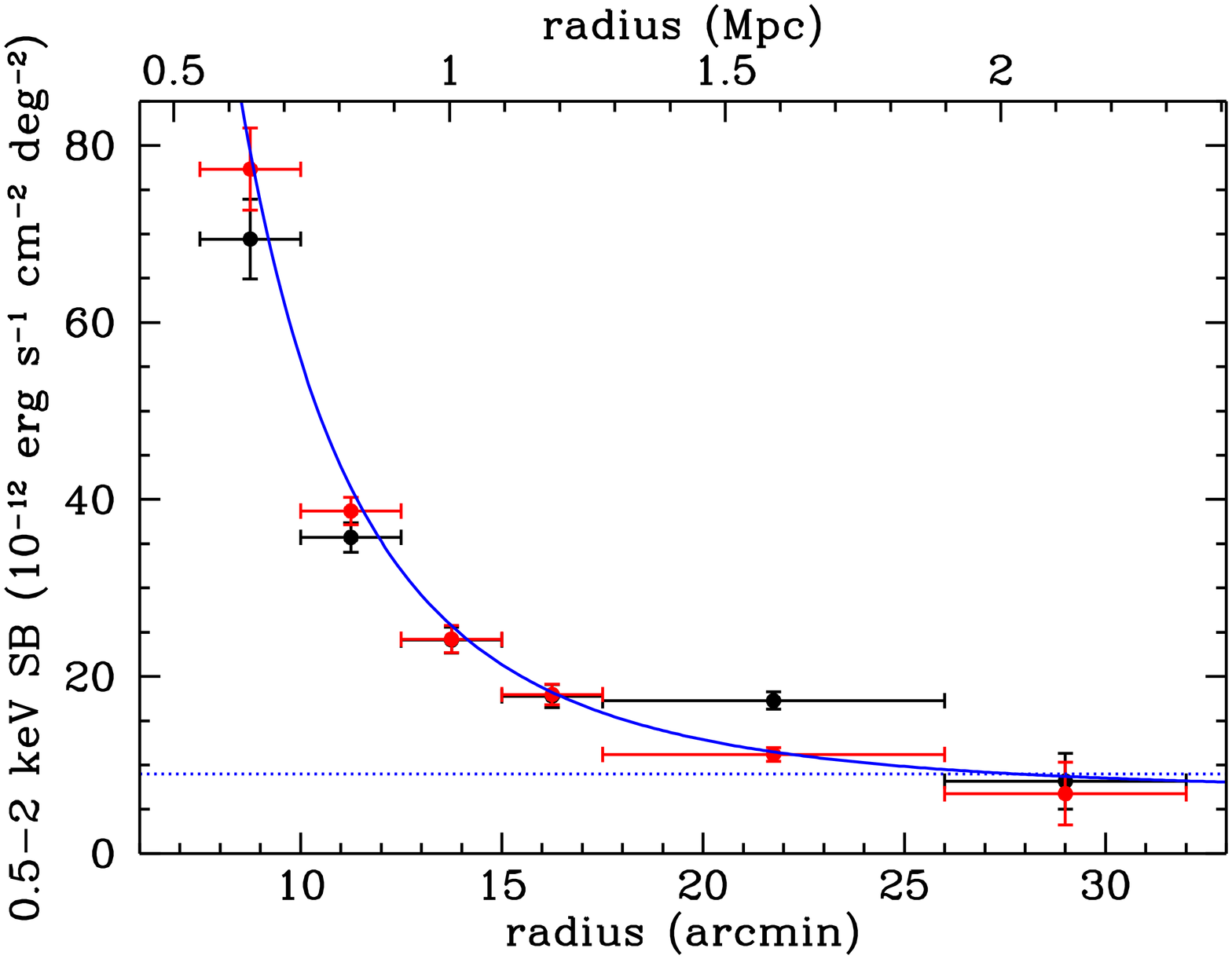}
\end{minipage}
\begin{minipage}{.24\textwidth}
\includegraphics[height=\textwidth,angle=0]{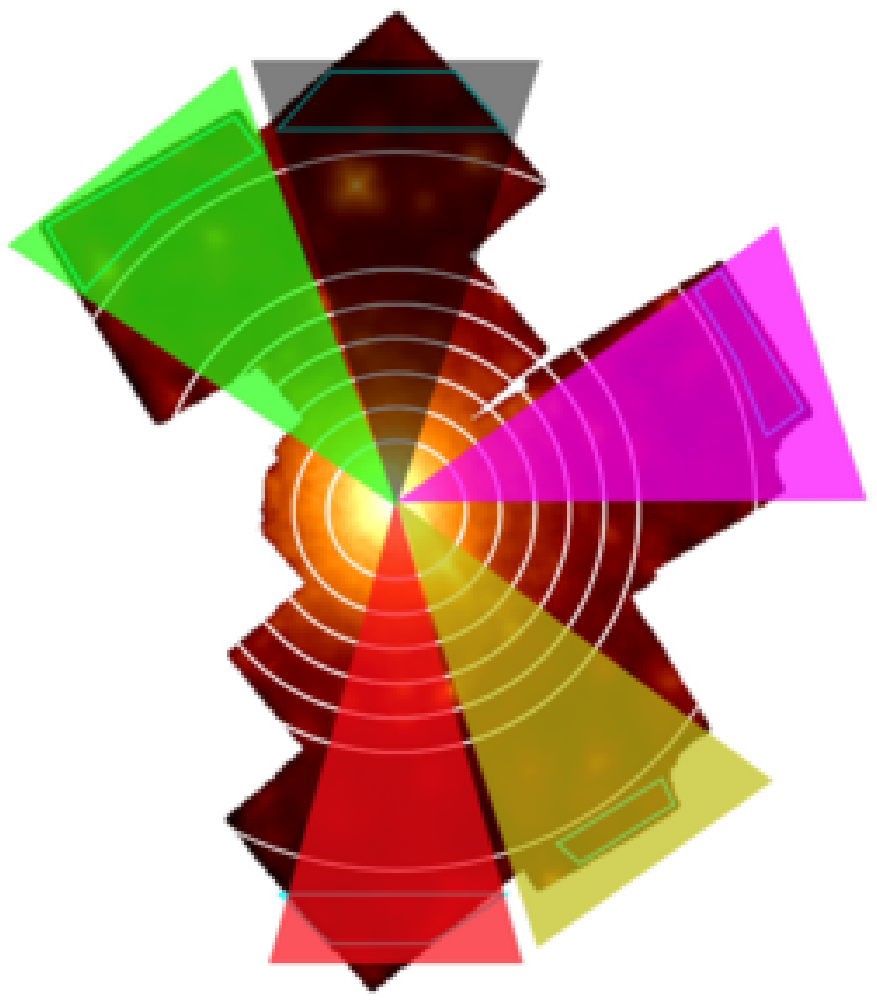}
\end{minipage}
\begin{minipage}{.38\textwidth}
\includegraphics[height=\textwidth,angle=270]{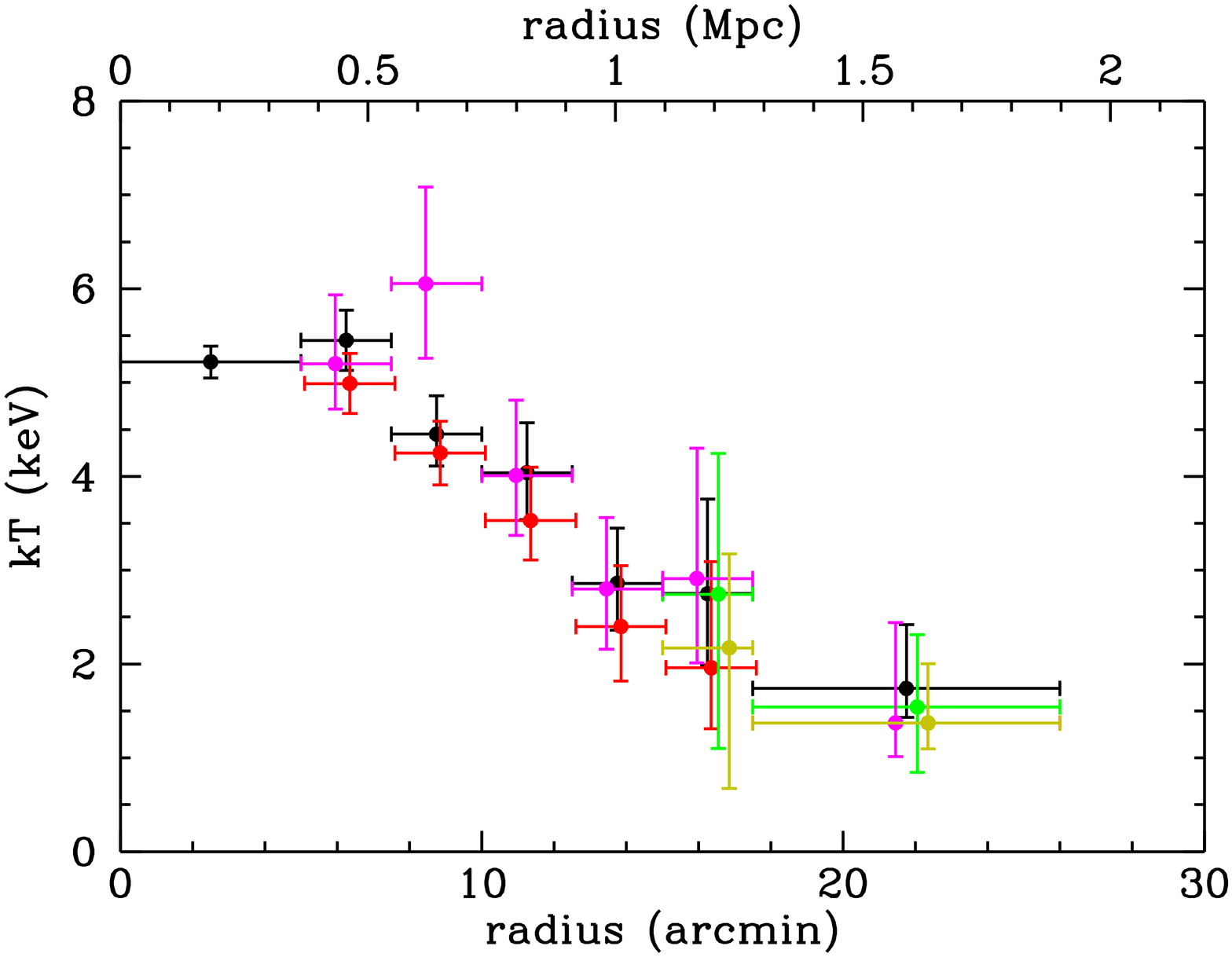}
\end{minipage}
\caption{Abell 1795 surface brightness (left) and temperature profiles
(right).  The image (center) shows color-coding for the radial directions
probed; for clarity, only the North and South directions are shown in the
surface brightness profile.  A 4-$\sigma$ difference is seen in the surface
brightness between the North and South \citep{Bautzetal2009}, and a rapid
decline in temperature is seen toward all directions.}
\label{fig:a1795tprof}
\end{figure}

Spatially resolved X-ray spectroscopy is one of the prime methods for
mapping the distribution of mass and intracluster plasma, and it is the
only way to measure heavy elements in the gas in galaxy clusters.  Although
\chandra\ and \xmm\ have successfully probed the central regions of
clusters \cite{Vikhlininetal2009b,Prattetal2007,Snowdenetal2008}, these
instruments have difficulty observing the faint emission near \rth\ due to
their relatively high and time-variable background, and in fact conditions
in the ICM beyond $\sim$0.5\rth\ have not been well studied by these
observatories.  \suzaku's low and stable background and substantial
effective area, however, have already enabled observations of a handful of
clusters that have successfully detected ICM emission out to and even
beyond \rth\
\cite{Georgeetal2009,Reiprichetal2009,Bautzetal2009,Hoshinoetal2010,Kawaharadaetal2010,Simionescuetal2011,Akamatsuetal2011}.
Our own \suzaku\ observations of Abell 1795 \cite{Bautzetal2009} find
significant deviations from hydrostatic equilibrium within \rtfh--\rth, in
a cluster which appears to be relaxed at smaller radii.  
These results
point to a rapidly declining temperature profile (see Figure
\ref{fig:a1795tprof}), an entropy profile that
is flatter than predicted for simple hierarchical structure formation
\cite{Voit2005}, and clearly asymmetric temperature and density profiles at
large radius.
The steep profile
may in turn result from cool, low-entropy infalling substructure, as
suggested by the simulations of 
\citet{Roncarellietal2006}.  Indeed, \citet{Burnsetal2010} successfully
reproduce with hydrodynamic simulations the temperature and entropy
profiles seen in Abell 1795, suggesting that bulk flows and turbulence from
accretion greatly affect the energetics of the outer ICM.

\section{Toward A Large Sample of Clusters with \suzaku}

While the previous results are intriguing, the \suzaku\ sample has been
small and it is unclear whether it represents clusters generally.
Starting in 2010, we have embarked on a comprehensive program
of observations of a large sample of clusters with \suzaku, utilizing a
recent \xmm\ cluster survey \cite{Snowdenetal2008}.  This survey is the
largest sample of galaxy clusters that have been probed to $\sim$ \rfh, and
it is the first sample large enough to find temperature profiles that seem
to differ significantly from theoretical predictions \cite{Juettetal2010}.
With \suzaku, we are able to capitalize on this work by extending the
radial coverage beyond that provided by \xmm\ to \rth, and provide robust,
independent checks of the \xmm\ results at smaller radii.  The goals of
this \suzaku\ project are to
(1) determine the temperature and density profiles to large radius for a
representative sample of apparently relaxed clusters, including those
discrepant with numerical simulations;
(2) search for azimuthal variations at large radius which may be 
indicative of the ongoing cluster accretion process; and
(3) provide a fiducial data set for detailed comparison with high
resolution numerical simulations.

To define our sample we first selected objects with high-quality \xmm\ data
to approximately \rfh.  This resulted in a sample with a variety of
temperature profiles that were falling, flat, or rising in the outer
regions.  The sample was further restricted using a numerical asymmetry
parameter to select clusters that appear relaxed in the \xmm\ images.  To
maximize \suzaku\ observing efficiency, we further selected clusters with 
$\rth \lesssim 17$ arcmin, so that the entire cluster volume within \rth\
could be observed in no more than four overlapping XIS pointings, while
allowing sufficient area beyond \rth\ for background estimation.  Clusters
were eliminated if they were too compact for the $\sim$ 2 arcmin \suzaku\
PSF ($\rth \lesssim 9$ arcmin), or if they were so centrally bright so that
the scattered light from the core would dominate cluster emission in the
\rfh--\rth\ region.  The current sample of 9 clusters is shown in Table
\ref{tab:sample}; it includes Abell 1795, which does not fully meet the
criteria outlined above, but has extensive \suzaku\ data from our previous
work.  We also include two clusters (Abell 1413 and Abell 2204) which have
single pointings from the archive; these data have been published
\citep{Reiprichetal2009,Hoshinoetal2010}.

\begin{figure}[t]
\begin{minipage}{.48\textwidth}
\includegraphics[width=\textwidth]{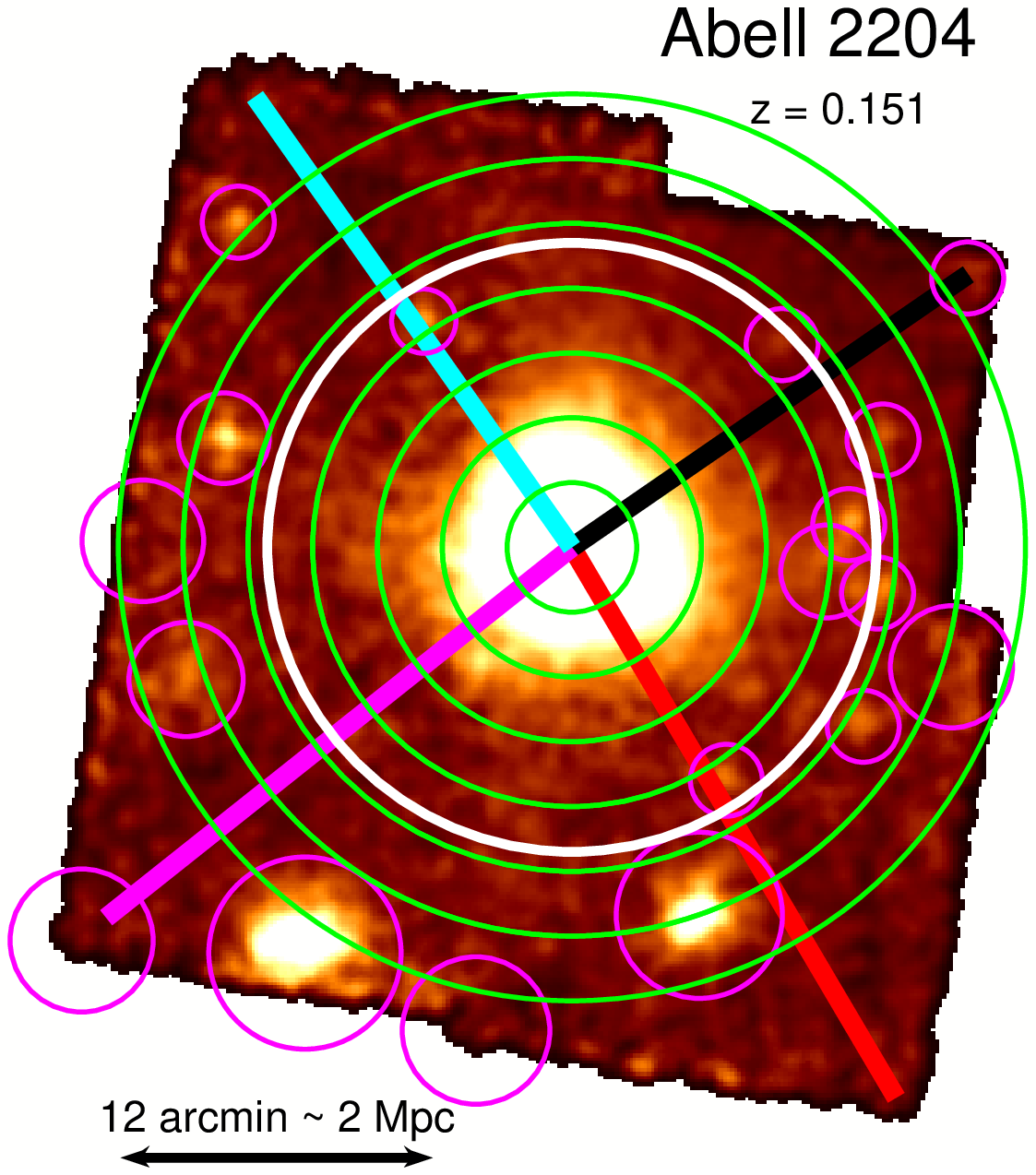}
\end{minipage}
\begin{minipage}{.48\textwidth}
\includegraphics[width=\textwidth]{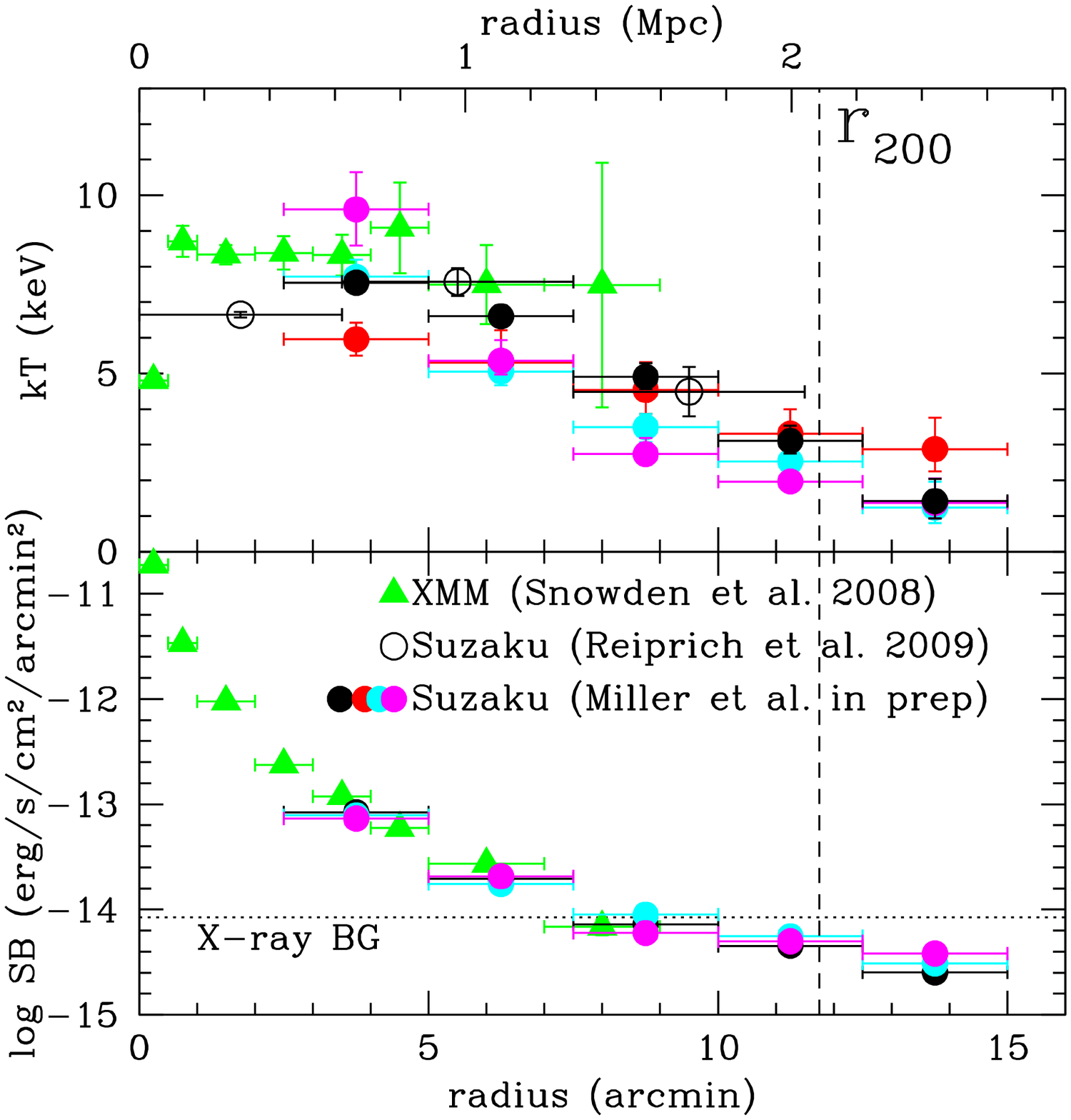}
\end{minipage}
\caption{Abell 2204 0.5--2 keV combined, exposure-corrected \suzaku\ image
(left) and profiles of temperature (right top) and surface brightness
(right bottom).  The profiles are measured in bins shown in the
image by the green annuli, along radial directions with arrows
corresponding to the colored points.  The white circle shows \rth = 2.3
Mpc.  Magenta circles identify point sources and other excised
features in the X-ray image.  The archival data from
\citet{Reiprichetal2009} are toward the northwest; those authors used
different annuli than we have in the current analysis.  The surface brightness
profile has been background-subtracted, so that we clearly see excess
emission beyond \rth; however, the uncertainty is larger than shown here,
as discussed in the text and in Figure \ref{fig:a2204sbimp}.}
\label{fig:a2204tprof}
\end{figure}

An example cluster is shown in Figure \ref{fig:a2204tprof}, with the
projected temperature and surface brightness profiles for Abell 2204
plotted along four radial directions.  The temperature drops off rapidly in
each direction, although there are differences from pointing to pointing.
Both the temperature and surface brightness are well-matched in the overlap
region to the \xmm\ results, which are azimuthally averaged.

\begin{table}[t]
\begin{tabular}{lccccc}
\hline
  \tablehead{1}{l}{b}{Cluster} 
& \tablehead{1}{c}{b}{$z$} 
& \tablehead{1}{c}{b}{$kT$\tablenote{Average $kT$ from \citep{Snowdenetal2008}}\\(keV)} 
& \tablehead{1}{c}{b}{\rth\\(arcmin)} 
& \tablehead{1}{c}{b}{$t_{\rm exp}$\tablenote{Total \suzaku\ exposure time from all pointings.}\\(ksec)} 
& \tablehead{1}{c}{b}{date obs.} \\
\hline
A 383      & 0.187 & \phn5.3 & \phn9.3 & 110 & July 2010 \\
A 773      & 0.216 & \phn7.2 & \phn9.5 & 200 & May 2011 \\
A 1068     & 0.147 & \phn4.7 &    10.8 & 200 & Oct 2011 \\
A 1413     & 0.135 & \phn7.6 &    14.8 & 170 & May 2010 + archive \\
A 1795     & 0.063 & \phn5.3 &    26.0 & 260 & June 2009 + archive \\
A 1914     & 0.174 &    11.5 &    14.5 & 160 & June 2010 \\
A 2204     & 0.151 & \phn8.6 &    11.8 & 140 & Sept 2010 + archive \\
A 2667     & 0.221 & \phn8.3 &    10.0 & 200 & July 2011 \\
RXCJ 0605  & 0.137 & \phn5.2 &    12.2 & 150 & May 2010 \\
\hline
\end{tabular}
\caption{Cluster sample}
\label{tab:sample}
\end{table}

\section{The Power of \suzaku, \xmm, and \chandra\ Together}
\label{sect:together}

\suzaku, \xmm, and \chandra\ together constitute an extremely powerful tool
for understanding the outskirts of clusters.  As noted above, \suzaku's low
background and substantial collecting area provide surface brightness
sensitivity unavailable with other instruments. \xmm\ and \chandra\ data
provide high angular resolution that resolves the shape of the temperature
profile near the cluster core, and excellent point source sensitivity in
the cluster outskirts to minimize the effects of fluctuations in the
background source density, as detailed below.  Finally, comparison of the
\xmm\ and \suzaku\ data in the regions of overlap provide important
cross-checks on systematic errors in temperature and flux
at the very low surface brightness levels we hope to study.

\begin{figure}[P]
\begin{minipage}{.33\textwidth}
\includegraphics[width=\textwidth]{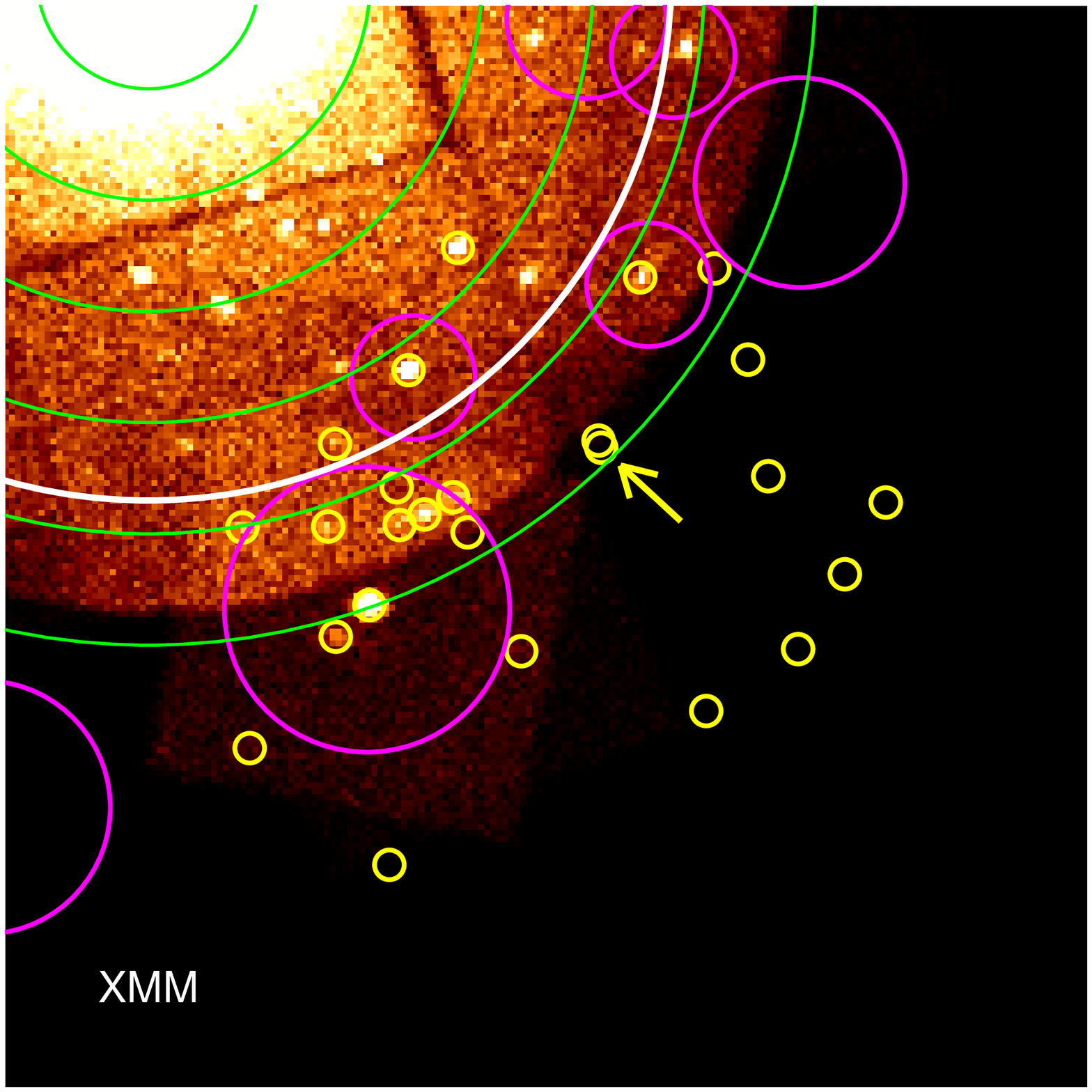}
\end{minipage}
\begin{minipage}{.33\textwidth}
\includegraphics[width=\textwidth]{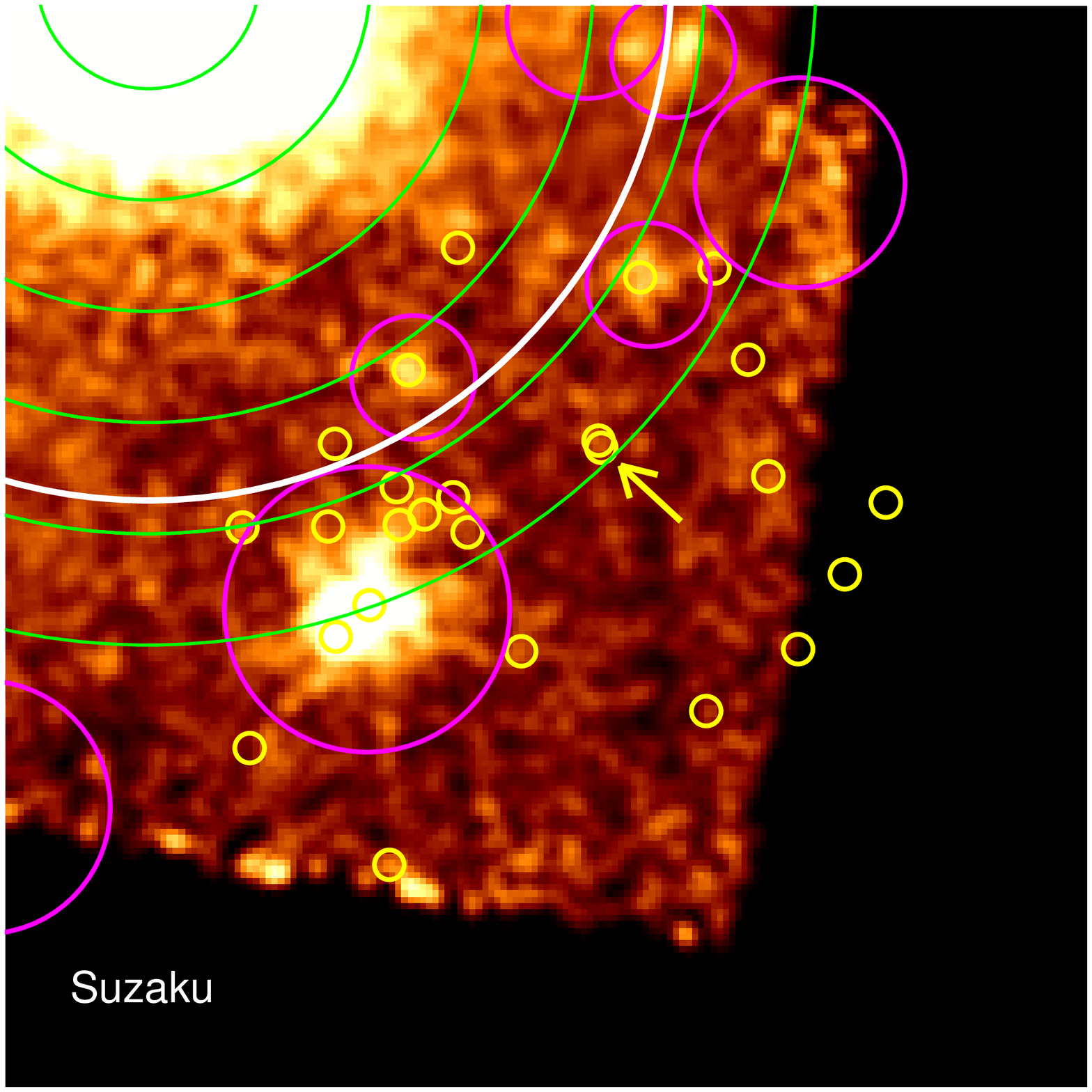}
\end{minipage}
\begin{minipage}{.33\textwidth}
\includegraphics[width=\textwidth]{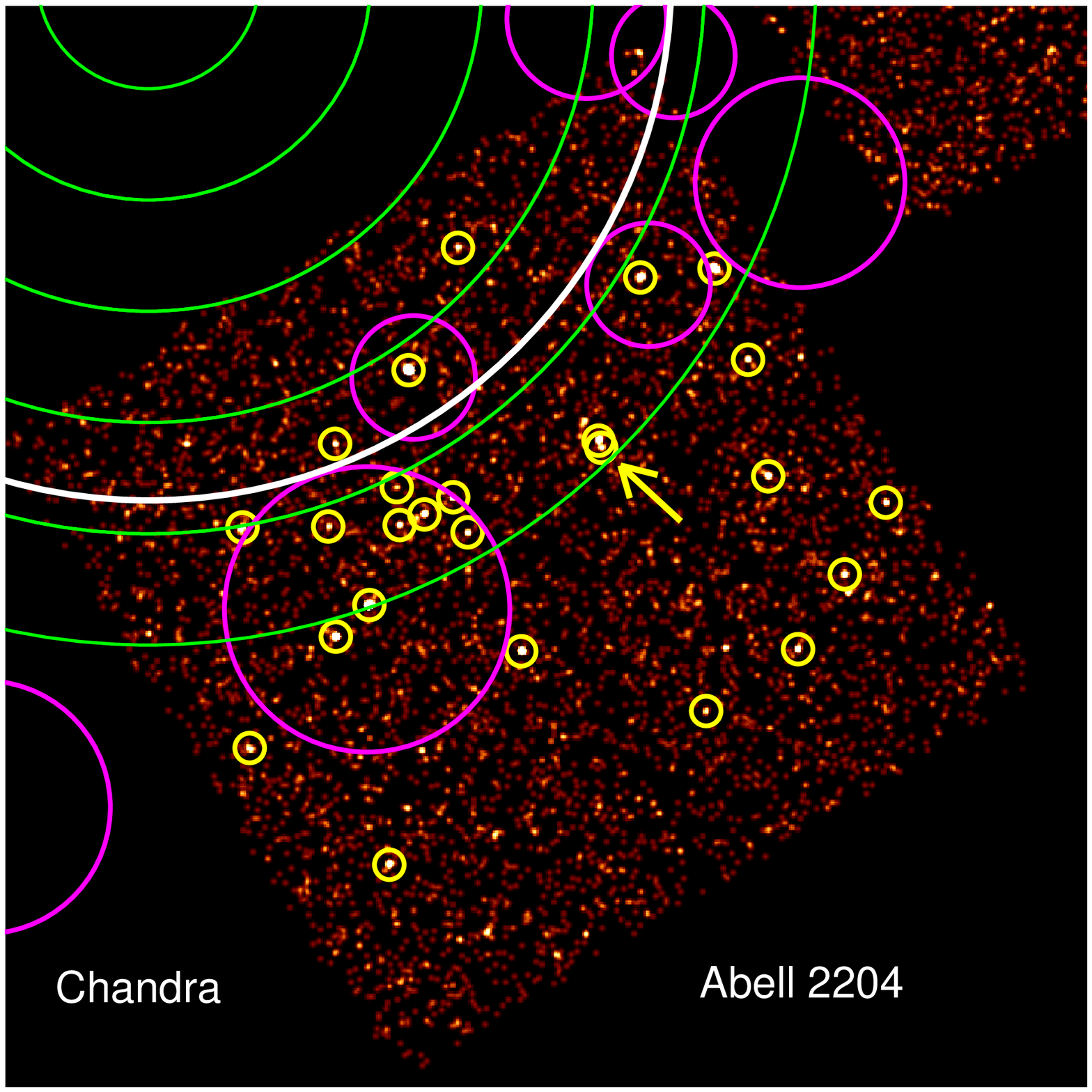}
\end{minipage}
\caption{Images of the southwest quadrant of Abell 2204 from \xmm, \suzaku,
and \chandra, showing detected point sources from \suzaku\ (magenta
circles) and \chandra\ (yellow circles).  Other notations are described
in Figure \ref{fig:a2204tprof}.  The resolving power of \chandra\ allows
detection of an order of magnitude more point sources than \suzaku, in an
exposure one-eighth as deep.  This depth allows tighter constraints on the
cosmic background uncertainty, as discussed in the text and illustrated in
Figure \ref{fig:a2204sbimp}.  The arrows indicate an example bright point
source, detectable by \chandra\ but not \suzaku, which dominates the
uncertainty in the outermost bin of Figure \ref{fig:a2204sbimp}.  Note that
the very bright point source has very similar flux in each image, however
due to the color scale and superior resolution it is less apparent in the
\chandra\ image.}
\label{fig:a2204pts}
\end{figure}

\begin{figure}[P]
\includegraphics[height=.8\textwidth,angle=270]{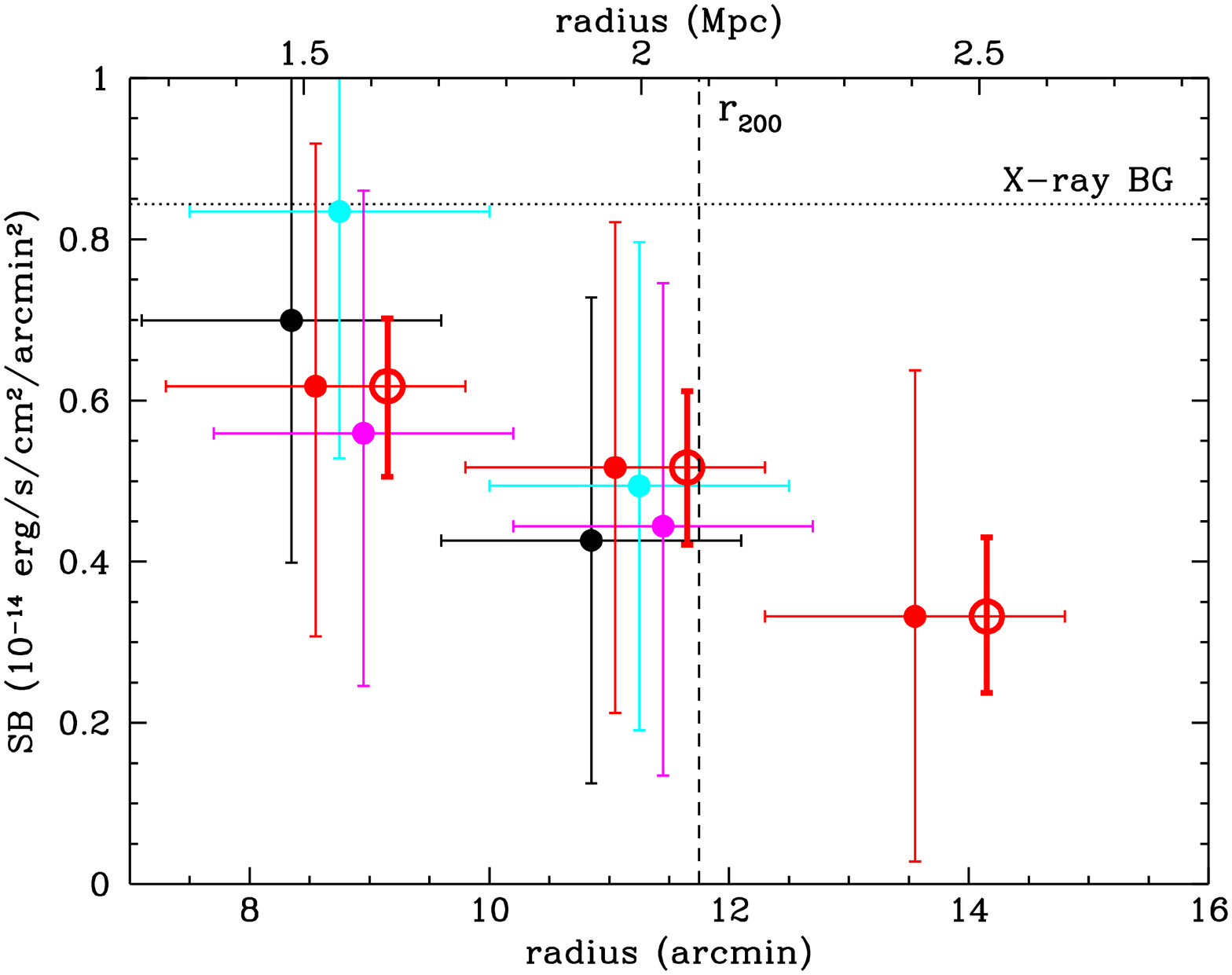}
\caption{Zoom-in of the Abell 2204 background-subtracted surface brightness
profile from Figure \ref{fig:a2204tprof}, showing only the outer three
radial bins and with the ordinate plotted on a linear scale.  The solid
points have error bars that include the uncertainty due to unresolved point
sources below the \suzaku\ flux limit.  The open points include error 
down to the \chandra\ point source flux limit for the single \suzaku\
pointing (shown in Figure \ref{fig:a2204pts}) with an overlapping \chandra\
snapshot.  The uncertainty is lower by a factor of $\sim$ 3.}
\label{fig:a2204sbimp}
\end{figure}

For small extraction regions ($\sim$ 0.01 deg$^{-2}$), fluctuations in the
unresolved source background dominate the surface brightness uncertainty.
For a typical 50 ksec exposure, we can expect to resolve and exclude
sources from the \suzaku\ data alone down to a flux of about \eez{-13}
\ergscms.  Assuming a differential source distribution along the lines of
\citet{Morettietal2003}, we expect surface brightness fluctuations in the
0.5--2 keV band to be $\sigma_B = 3.9\eex{-12} \Omega_{0.01}^{-1/2}$
\ergscmsdeg, where $\Omega_{0.01}$ is the region solid angle in units of
0.01 deg$^{-2}$ \citep{Bautzetal2009}.  This is 30\% of the typical X-ray
background level measured in our Abell 1795 \suzaku\ background regions,
and it accounts for 40\% of the detected cluster emission at \rfh--\rth\
toward the north in Abell 1795 \citep{Bautzetal2009}.
Including high spatial resolution data from \xmm\ or \chandra\ allows us to
remove fainter sources, lowering our point source exclusion threshold by a
factor of 10 in flux and reducing $\sigma_B$ by a factor of $\sim$ 3--4.
While increasing the solid angle will also mitigate fluctuations due to
cosmic variance, we wish to trace azimuthal variations.  This requires
restricting extraction regions to at most a few times 0.01 deg$^{-2}$ for
the less extended clusters, and therefore we have required our \suzaku\
pointings to overlap as much as possible with available \xmm\ or \chandra\
data so that we can adequately eliminate or model point sources.  

These clusters generally lack \xmm\ coverage in the outskirts, so we have
begun a parallel project to obtain \chandra\ snapshot observations to fill
in these regions.  An example is illustrated in Figures \ref{fig:a2204pts}
and \ref{fig:a2204sbimp}.  Including 5 ksec \chandra\ snapshots increases
the number of detected point sources by an order of magnitude and reduces
our surface brightness uncertainty by a factor of $\sim$ 3.  For a
given detected \chandra\ point source, we either exclude this source 
from the \suzaku\ extraction or include it as an additional component in
the background fit, depending on the point source flux and size of the full
extraction region.  About one quarter of the \chandra\ observations are
completed, and this analysis remains ongoing.

\section{A Case Study: The Outskirts of RXCJ 0605}

\begin{figure}[p]
\begin{minipage}{.5\textwidth}
\includegraphics[width=\textwidth]{tprof_wimg.eps}
\end{minipage}
\begin{minipage}{.5\textwidth}
\includegraphics[width=\textwidth]{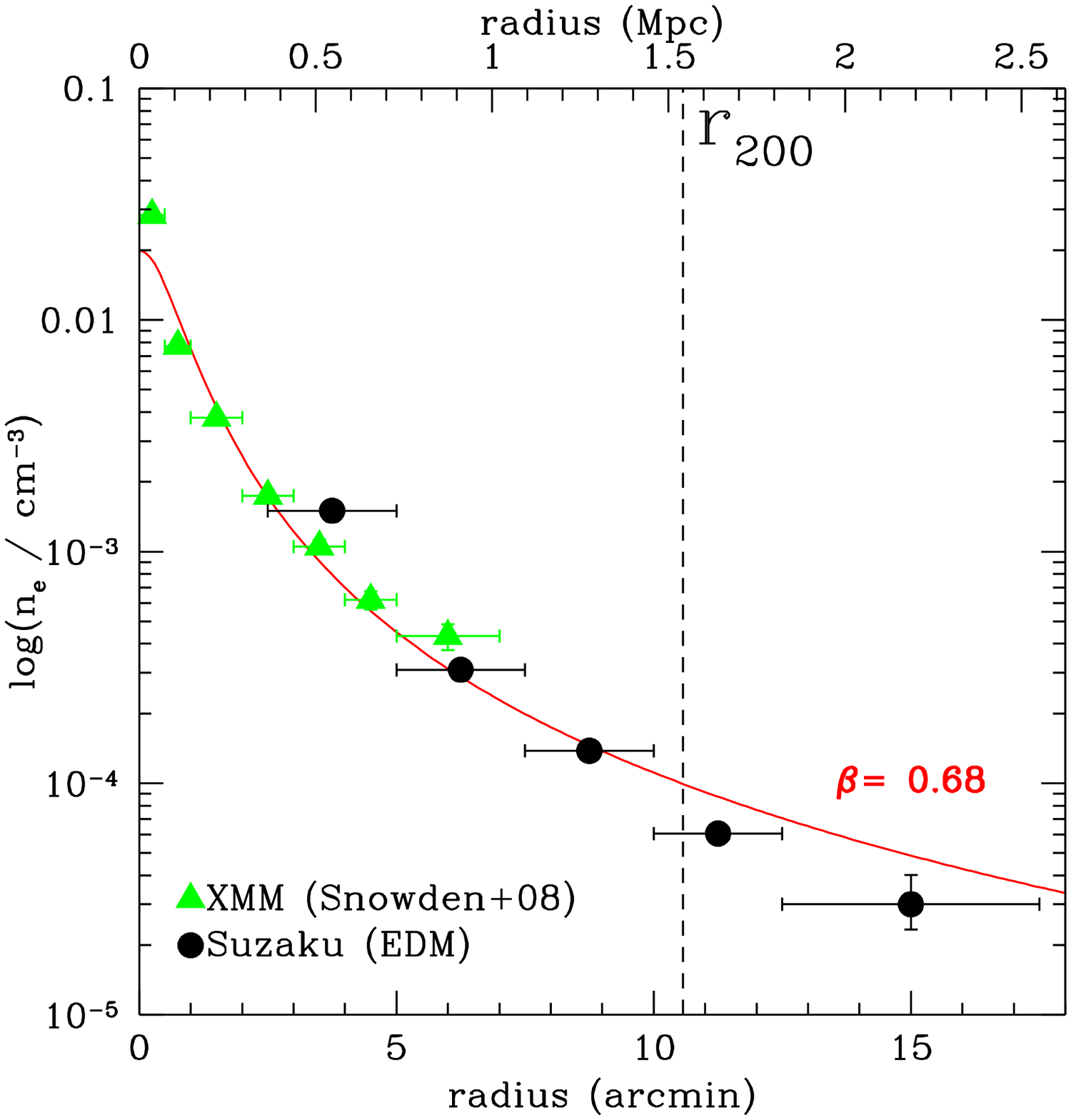}
\end{minipage}
\caption{RXCJ 0605 temperature and electron density profiles.  The inlay
is the 0.5--2 keV combined, exposure-corrected \suzaku/XIS image,
with the same notations as in Figure \ref{fig:a2204tprof}.}
\label{fig:temp_ne}
\end{figure}

\begin{figure}[p]
\begin{minipage}{.5\textwidth}
\includegraphics[width=\textwidth]{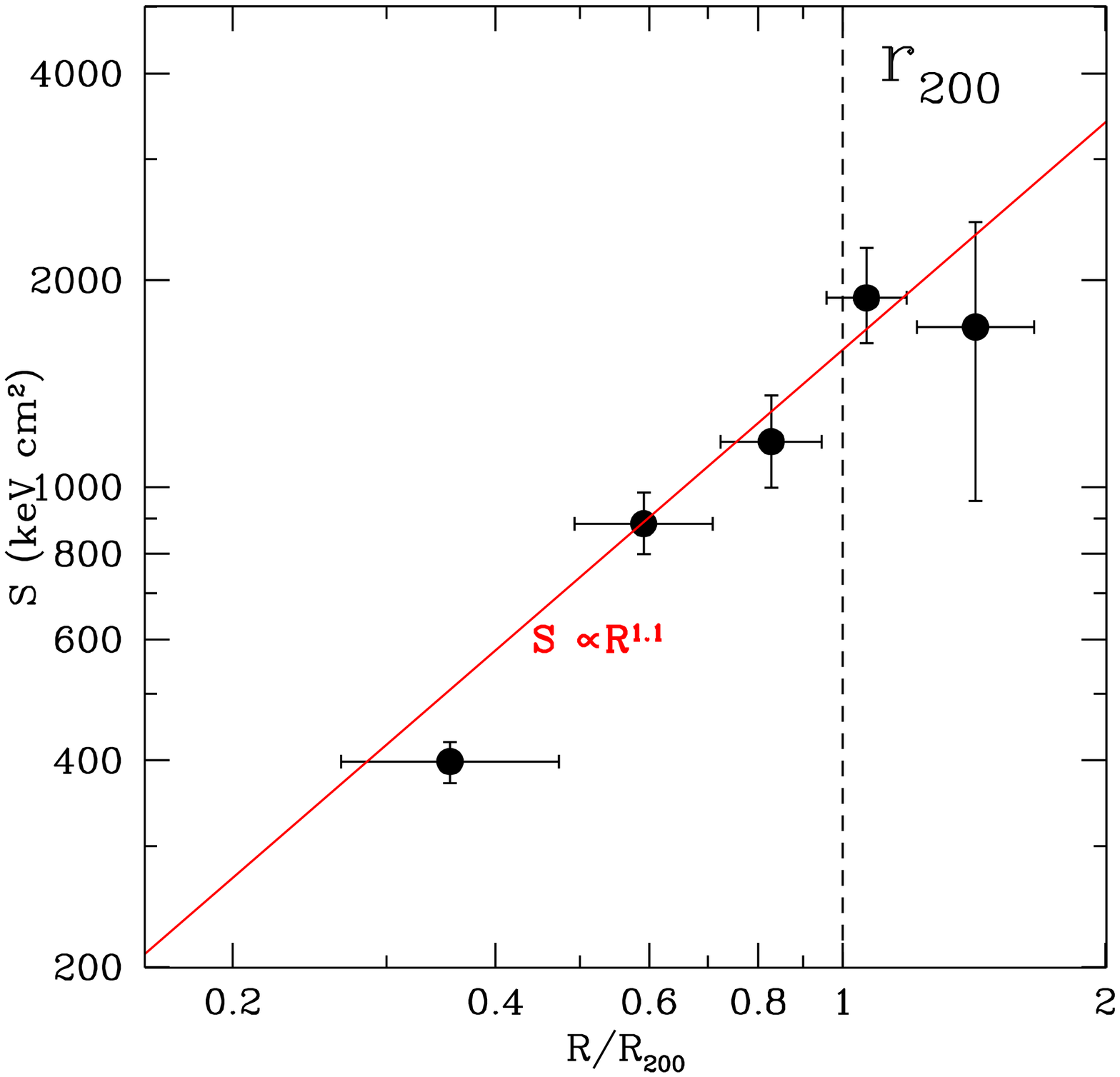}
\end{minipage}
\begin{minipage}{.5\textwidth}
\includegraphics[width=\textwidth]{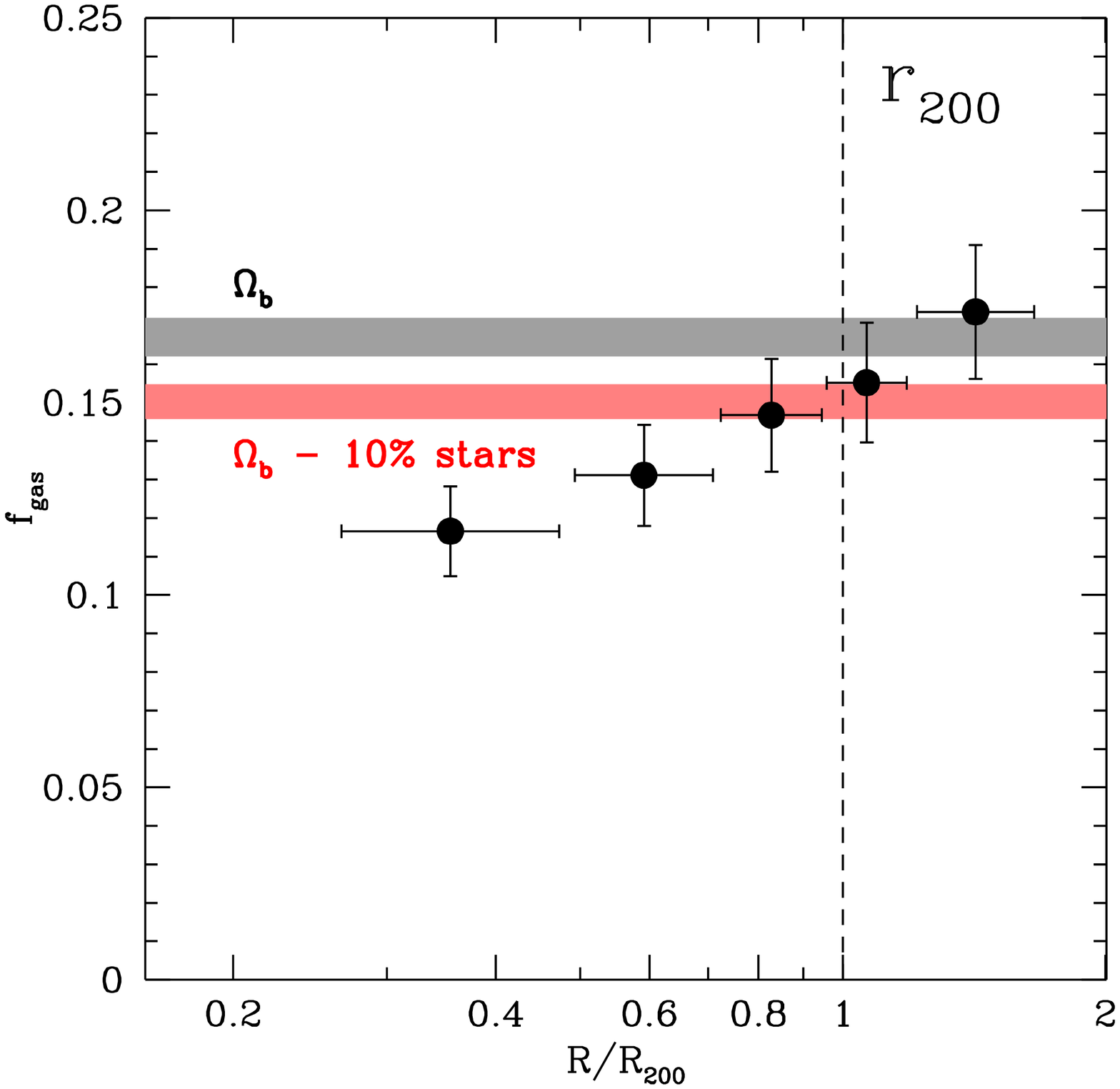}
\end{minipage}
\caption{RXCJ 0605 entropy and \fgas\ profiles.  The $S \propto R^{1.1}$
relation is from \citet{Voit2005}.  The shaded regions in the \fgas\ plot
show current estimates of the cosmic baryon fraction.}
\label{fig:ent_fgas}
\end{figure}

The cluster RXCJ 0605.8-3518 (Abell 3378) appears to have a rising $kT$
profile to \rfh\ in the \xmm\ analysis \citep{Snowdenetal2008}.  We present
azimuthally-averaged, deprojected gas profiles from our \suzaku\
observations in Figures \ref{fig:temp_ne} and \ref{fig:ent_fgas}.  The
temperature decreases beyond the cluster core, as does the electron
density, which is consistent with the \xmm\ result in the overlap region
and a $\beta = 0.68$ profile out well beyond \rth.  The entropy profile
increases steadily to as far as can be measured, in line with theoretical
predictions \citep{Voit2005} and at odds with results from previous
clusters \citep[e.g., Abell 1795;][]{Bautzetal2009}.  The gas fraction is
fully consistent with the cosmic baryon fraction at large radius.

These results are also at odds with a recent study of the Perseus cluster
\citep[][and elsewhere in these proceedings]{Simionescuetal2011}, which
finds a flattening of the entropy profile (similar to other previous
cluster studies) and an observed gas fraction well in excess of the baryon
fraction near \rth.  Such discrepancies with predictions can be explained
by clumping of the ICM gas; since the X-ray emissivity is proportional to
the square of the gas density, small dense regions will dominate the
emission, biasing the entropy lower and the gas fraction higher compared to
extrapolation from the more homogeneous inner regions.  At this point, the
importance of gas clumping near the virial radius remains unclear.  The
results from the Perseus cluster are clearly inconsistent with a
homogeneous ICM in hydrostatic equilibrium, and clumping neatly explains
the discrepancies between observations and predictions in the thermodynamic
gas profiles.  Clumping at various scales is also predicted by numerical
simulations \citep{Roncarellietal2006,Burnsetal2010,NagaiLau2011}.  Yet our
initial results and a recent \suzaku\ study of the fossil group RXJ
1159+5531 \citep{Humphreyetal2011} are consistent with expectations from a
homogeneous ICM and need not invoke clumping.  To what extent environment, 
observing location (e.g., along a filament), or instrumental effects 
might play a role in these different results is an open question.  Results
from our full sample of clusters, as well as additional on-going studies of
Perseus, Virgo, and other nearby bright clusters, will help solve this
question.

\clearpage


\begin{theacknowledgments}
We thank both the organizers and attendees of ``\suzaku\ 2011'' for a
highly enlightening and enjoyable meeting.  This work was supported by NASA
grants NNX10AV02G and NNX09AE58G, and by SAO grant GO1-12165X.  
\end{theacknowledgments}



\bibliographystyle{aipproc}   


\begin{thebibliography}{22}
\expandafter\ifx\csname natexlab\endcsname\relax\def\natexlab#1{#1}\fi
\providecommand{\enquote}[1]{``#1''}
\expandafter\ifx\csname url\endcsname\relax
  \def\url#1{\texttt{#1}}\fi
\expandafter\ifx\csname urlprefix\endcsname\relax\def\urlprefix{URL }\fi
\providecommand{\eprint}[2][]{\url{#2}}

\bibitem[{Evrard} et~al.(1996)]{Evrardetal1996}
A.~E. {Evrard}, C.~A. {Metzler}, and J.~F. {Navarro}, \emph{\apj} \textbf{469},
  494 (1996)

\bibitem[{Voit}(2005)]{Voit2005}
G.~M. {Voit}, \emph{Reviews of Modern Physics} \textbf{77}, 207 (2005)

\bibitem[{Roncarelli} et~al.(2006)]{Roncarellietal2006}
M.~{Roncarelli}, S.~{Ettori}, K.~{Dolag}, L.~{Moscardini}, S.~{Borgani}, and
  G.~{Murante}, \emph{\mnras} \textbf{373}, 1339 (2006)

\bibitem[{Burns} et~al.(2010)]{Burnsetal2010}
J.~O. {Burns}, S.~W. {Skillman}, and B.~W. {O'Shea}, \emph{\apj} \textbf{721},
  110 (2010)

\bibitem[{Navarro} et~al.(1997)]{NFW1997}
J.~F. {Navarro}, C.~S. {Frenk}, and S.~D.~M. {White}, \emph{\apj} \textbf{490},
  493 (1997)

\bibitem[{Borgani} et~al.(2004)]{Borganietal2004}
S.~{Borgani}, {et al.}, \emph{\mnras} \textbf{348}, 1078 (2004)

\bibitem[{Vikhlinin} et~al.(2009{\natexlab{a}})]{Vikhlininetal2009a}
A.~{Vikhlinin}, {et al.}, \emph{\apj} \textbf{692}, 1033 (2009{\natexlab{a}})

\bibitem[{Allen} et~al.(2004)]{Allenetal2004}
S.~W. {Allen}, R.~W. {Schmidt}, H.~{Ebeling}, A.~C. {Fabian}, and L.~{van
  Speybroeck}, \emph{\mnras} \textbf{353}, 457 (2004)

\bibitem[{Bautz} et~al.(2009)]{Bautzetal2009}
M.~W. {Bautz}, {et al.}, \emph{\pasj} \textbf{61}, 1117 (2009)

\bibitem[{Vikhlinin} et~al.(2009{\natexlab{b}})]{Vikhlininetal2009b}
A.~{Vikhlinin}, {et al.}, \emph{\apj} \textbf{692}, 1060 (2009{\natexlab{b}})

\bibitem[{Pratt} et~al.(2007)]{Prattetal2007}
G.~W. {Pratt}, H.~{B{\"o}hringer}, J.~H. {Croston}, M.~{Arnaud}, S.~{Borgani},
  A.~{Finoguenov}, and R.~F. {Temple}, \emph{\aap} \textbf{461}, 71 (2007)

\bibitem[{Snowden} et~al.(2008)]{Snowdenetal2008}
S.~L. {Snowden}, R.~F. {Mushotzky}, K.~D. {Kuntz}, and D.~S. {Davis},
  \emph{\aap} \textbf{478}, 615 (2008)

\bibitem[{George} et~al.(2009)]{Georgeetal2009}
M.~R. {George}, A.~C. {Fabian}, J.~S. {Sanders}, A.~J. {Young}, and H.~R.
  {Russell}, \emph{\mnras} \textbf{395}, 657 (2009)

\bibitem[{Reiprich} et~al.(2009)]{Reiprichetal2009}
T.~H. {Reiprich}, {et al.}, \emph{\aap} \textbf{501}, 899 (2009)

\bibitem[{Hoshino} et~al.(2010)]{Hoshinoetal2010}
A.~{Hoshino}, {et al.}, \emph{\pasj} \textbf{62}, 371 (2010)

\bibitem[{Kawaharada} et~al.(2010)]{Kawaharadaetal2010}
M.~{Kawaharada}, {et al.}, \emph{\apj} \textbf{714}, 423 (2010)

\bibitem[{Simionescu} et~al.(2011)]{Simionescuetal2011}
A.~{Simionescu}, {et al.}, \emph{Science} \textbf{331}, 1576 (2011)

\bibitem[{Akamatsu} et~al.(2011)]{Akamatsuetal2011}
H.~{Akamatsu}, A.~{Hoshino}, Y.~{Ishisaki}, T.~{Ohashi}, K.~{Sato}, Y.~{Takei},
  and N.~{Ota}, \emph{PASJ}, submitted (2011), \eprint{astro-ph/1106.5653}

\bibitem[{Juett} et~al.(2010)]{Juettetal2010}
A.~M. {Juett}, D.~S. {Davis}, and R.~{Mushotzky}, \emph{\apjl} \textbf{709},
  L103 (2010)

\bibitem[{Moretti} et~al.(2003)]{Morettietal2003}
A.~{Moretti}, S.~{Campana}, D.~{Lazzati}, and G.~{Tagliaferri}, \emph{\apj}
  \textbf{588}, 696 (2003)

\bibitem[{Nagai} and {Lau}(2011)]{NagaiLau2011}
D.~{Nagai}, and E.~T. {Lau}, \emph{\apjl} \textbf{731}, L10 (2011)

\bibitem[{Humphrey} et~al.(2011)]{Humphreyetal2011}
P.~J. {Humphrey}, D.~A. {Buote}, F.~{Brighenti}, H.~M.~L.~G. {Flohic},
  F.~{Gastaldello}, and W.~G. {Mathews}, \emph{ApJ}, submitted  (2011),
  \eprint{astro-ph/1106.3322}

\end{thebibliography}

\end{document}